\pdfoutput=1
\documentclass[sigconf]{acmart} 

\usepackage{array}
\usepackage{xspace}
\newcommand{\ie}{{i.e.,}\xspace}
\newcommand{\eg}{{e.g.,}\xspace}

\newcommand{\ea}{{et~al.}\xspace}

\newcommand{\etc}{{etc.}\xspace}
\newcommand{\sectionref}[1]{{Section}~\ref{#1}}
\newcommand{\sectref}[1]{{Sect.}~\ref{#1}}
\newcommand{\bstart}[1]{\vspace{1mm} \noindent{\textbf{#1:}}}
\newcommand{\bstartnc}[1]{\vspace{1mm} \noindent{\textbf{#1}}}

\newcounter{ci}
\definecolor{indexBlue}{cmyk}{0.9,0.8,0,0}
\definecolor{darkBlue}{cmyk}{0.8,0.5,0,0.3}
\newcommand{\cons}[2]{\refstepcounter{ci}\label{#2}\bstart{\textcolor{indexBlue}{(C\theci)} #1}}
\newcommand{\ci}[1]{\textcolor{indexBlue}{\textbf{C\ref{#1}}}}

\AtBeginDocument{%
  \providecommand\BibTeX{{%
    \normalfont B\kern-0.5em{\scshape i\kern-0.25em b}\kern-0.8em\TeX}}}

\copyrightyear{2022}
\acmYear{2022}
\setcopyright{acmcopyright}\acmConference[CHI '22]{CHI Conference on Human Factors in Computing Systems}{April 29-May 5, 2022}{New Orleans, LA, USA}
\acmBooktitle{CHI Conference on Human Factors in Computing Systems (CHI '22), April 29-May 5, 2022, New Orleans, LA, USA}
\acmPrice{15.00}
\acmDOI{10.1145/3491102.3517484}
\acmISBN{978-1-4503-9157-3/22/04} 

\begin{document}

\title{Personalization Trade-offs in Designing a Dialogue-based Information System for Support-Seeking of Sexual Violence Survivors}

\author{Hyeok Kim}
\affiliation{%
  \institution{Northwestern University}
  \city{Evanston}
  \state{IL}
  \country{U.S.A.}}
\email{hyeokkim2024@u.northwestern.edu}

\author{Youjin Hwang}
\affiliation{%
  \institution{Seoul National University}
  \city{Seoul}
  \country{Korea}}
\email{youjin.h@snu.ac.kr}

\author{Jieun Lee}
\affiliation{%
  \institution{Seoul National University}
  \city{Seoul}
  \country{Korea}}
\email{leex7164@snu.ac.kr}

\author{Youngjin Kwon}
\affiliation{%
  \institution{Seoul National University}
  \city{Seoul}
  \country{Korea}}
\email{shb02192@snu.ac.kr}

\author{Yujin Park}
\affiliation{%
  \institution{Seoul National University}
  \city{Seoul}
  \country{Korea}}
\email{yujin607@snu.ac.kr}

\author{Joonhwan Lee}
\authornote{Corresponding author}
\affiliation{%
  \institution{Seoul National University}
  \city{Seoul}
  \country{Korea}}
\email{joonhwan@snu.ac.kr}

\renewcommand{\shortauthors}{Kim et al.}

\begin{abstract}
The lack of reliable, personalized information often complicates sexual violence survivors' support-seeking.
Recently, there is an emerging approach to conversational information systems for support-seeking of sexual violence survivors, featuring personalization with wide availability and anonymity.
However, a single best solution might not exist as sexual violence survivors have different needs and purposes in seeking support channels.
To better envision conversational support-seeking systems for sexual violence survivors, we explore personalization trade-offs in designing such information systems.
We implement a high-fidelity prototype dialogue-based information system through four design workshop sessions with three professional caregivers and interviewed with four self-identified survivors using our prototype.
We then identify two forms of personalization trade-offs for conversational support-seeking systems: (1) specificity and sensitivity in understanding users and (2) relevancy and inclusiveness in providing information.
To handle these trade-offs, we propose a \textit{reversed approach} that starts from designing information and \textit{inclusive tailoring} that considers unspecified needs, respectively.
\end{abstract} 



\maketitle

\section{Introduction}\label{sec:intro}
Sexual violence (or, sexual assault, abuse) generally refers to a coerced sexual act upon an individual without consent~\cite{rainn}, including (but not limited to) rape, molestation, and incest. 
Sexual violence survivors often suffer from short-term and/or long-term physical~\cite{campbell2002health,tavara2006sexual} and/or mental~\cite{siegel1990reactions,au2013co,burnam1988sexual} problems. 
They are often discouraged from seeking support due to self-disclosure~\cite{quadara2008responding}, acquainted perpetrators~\cite{millar2002immediate}, fear of revenge~\cite{justice2013}, financial insecurity~\cite{salisbury1986counseling}, and stereotypes like rape myths~\cite{fanflik2007victim}.
Above all, the lack of personalized, reliable information about ``access'' to professional support (\eg~medical checkup, legal assistance, \etc) adds another barrier for survivors to request the support they need~\cite{logan2005barriers,walsh2010disclosure,clark2010justice}.

Recently, there is an emerging dialogue-based approach to personalized informational assistance for sexual violence survivors (\eg~Gabbie~\cite{gabbie}, Spot~\cite{spot}, Hello Cass~\cite{cass}, Bauer~\ea~\cite{bauer2019}, Park~and~Lee~\cite{park2021}), presumably because conversational methods afford anonymity, promptness, wide accessibility (\eg~\cite{Heerden2017,wang2018}), and social interaction (\eg~\cite{woebot,wysa}).
Park~and~Lee~\cite{park2021} suggest dialogue-based interactions can help reduce various burdens that sexual violence survivors may experience during their support-seeking process, such as amount of time spent, privacy, emotional concerns.
In particular, they indicate that conversational agents can reduce `availability burden' (\ie~perceived lack of sufficient resources) through personalizing available resources.
However, sexual violence survivors are also individuals with different needs and purposes (\eg~medical treatment, legal support, and sense-making, \etc), so designers should make various design decisions according to their target user characteristics (\eg~informational needs, preferred conversation methods) as well as addressing practical constraints (\eg~data storage, statute of limitations). 

While prior work~\cite{park2021,park2020can,maeng2021,ahn2020chatbot,freed2018stalker} informs about various aspects around interface design of information systems for sexual violence survivors, deeper discussion on \textit{how to design personalization} (\ie~how to ask personalizing questions and personalize information for users) often lacks.
Therefore, understanding design trade-offs around personalization is a key step toward better reasoning about information systems for different types of users.
We aim at identifying major design trade-offs that arise due to personalization of such conversational information systems and propose potential strategies to address those trade-offs to motivate future collaboration of designers, practitioners, and end users.

To do so, we first implement a prototype system through four design workshop sessions with three sexual assault support professionals and conduct preliminary evaluation through two focus group interviews.
Our prototype asks questions about users' cases, classify users into groups, and provide personalized information. 
Our prototype is designed to be used for a short duration to encourage survivors to promptly access expert cares, and it identifies users in urgent cases based on the practice guideline to inform about immediate actions they may need to take.
To address privacy concerns of survivors, our prototype does not include their direct identification information (\eg~email, name).

Using this prototype, we conducted an interview study with four self-identified sexual violence survivors to understand how users would perceive our design rationales. 
We describe interview results in terms of comfortability, relevancy, difficulty, and actionable instructions.
While the interviewees evaluated our prototype as useful and correct in general, they suggested potential improvements, such as using easier expressions, reordering conversation in an ascending order of sensitivity, and being accountable for the questions our prototype asks.
Our interviewees showed bipartite opinions in terms of relevancy and sensitivity; some of them found it as quite relatable, but the others found somewhat less related to them. 
They also differently perceived questions of our prototype as necessary for personalization or burdening to answer.
Our interviewees found actionable instructions provided by our prototype useful by envisioning what they might experience after contacting a support institution.

Based on the findings from our design workshop and user study, we discuss two types of important trade-offs in terms of personalization and propose design strategies to handle them.
First, there is a trade-off between specificity of questions for personalization and sensitivity that such specifying questions can cause.
To handle this trade-off, we take a \textit{reversed approach} that first considers the outcome of interaction (provided information) and decides specific questions to ask (\ie~reversed to the users' journey).
Second, we observe a trade-off between specificity in finding emergent and particular needs of users and inclusiveness in addressing emerging needs that users develop as they become more aware of their situation.
We cope with this trade-off through an \textit{inclusive tailoring} approach comprised of low-level techniques like clarification, screening, ordering, and a response option for `not sure.'
We discuss implication of our methods and findings for future support systems for support-seeking of sexual violence survivors.
\section{Background}\label{sec:background}

\subsection{Desired Qualities for Support-seeking Systems for Sexual Violence Survivors}\label{sec:support}
Empirical findings from HCI, justice, and social science indicate that in seeking professional support (\eg~healthcare, counseling, reporting, \etc), sexual violence survivors often need to search across a vast amount of support resources to find optimal solutions for themselves while minimizing any potential risks like disclosure and ensuring validity of information they found.
Recently, Park~and~Lee~\cite{park2021} identify various burdens of sexual violence survivors in seeking support, including privacy, amount of time, data leaks, cognitive resources, emotional burdens, and financial difficulties.
Below, we describe anonymity, wide accessibility, promptness, correctness, and personalization as desired qualities of support-seeking systems for sexual violence survivors to be useful.

\subsubsection{Needs for Wide accessibility, anonymity, and promptness}
As survivors may have difficulty in accessing professional support due to limited channels and privacy concerns, wide accessibility, anonymity, and promptness are necessary for support-seeking systems.
Right after experiencing sexual assault, people are less likely to know appropriate information about support channels~\cite{logan2005barriers,walsh2010disclosure} often due to insufficient socioeconomic resources~\cite{cattaneo2008sexual,henning2002utilization}.
Accessing and using professional support may impose burden of disclosure on survivors~\cite{quadara2008responding,justice2013}, which in turn discourages them from seeking information~\cite{salisbury1986counseling}.
As anonymity on online platforms makes it more comfortable to disclose their experiences~\cite{andalibi2018social}, some sexual violence survivors make disclosure on online communities for sense-making and information-seeking~\cite{andalibi2016understanding,razi2020let}, but seeking information from online forums can be ignored or delayed~\cite{zhang2007qume,weinberg1995computer}.
Under a highly patriarchal society~\cite{naseem2020designing} or a threat by intimate partner abuser~\cite{freed2018stalker,matthews2017stories}, wider and anonymous access has much higher priority because family members or abusers can access survivors' online activity or physical device. 
For example, if a system requires many steps like user registration and/or disclosure of private information, users have to leave more digital records on their device which abusers could detect.
Furthermore, considering that many survivors seek professional support after physical injuries or life threats \cite{ullman2001correlates,fanslow2010help}, it is crucial for them to access those support in a timely manner.
Ahmed~\ea~\cite{ahmed2014protibadi} propose instant emission of rescue signal to trusted contacts to empower survivors by enhancing the possibility of prompt care.

\subsubsection{Needs for correct, personalized information}
On top of the desired qualities described above, correct information is important to better support sexual violence survivors.
For example, empirical studies find that information from online platforms can be often invalid~\cite{westbrook2007digital,gustafson2008internet} because its providers are not likely to be professional.
Analyzing posts of support-seeking for sexual experiences on online platforms, Razi~\ea~\cite{razi2020let} stress  the needs for professional moderation to guarantee the quality of online information. 
Given that sexual assault involves highly personal narratives, a key quality for providing correct information is personalization.
Survivors may find instructions or other people's experiences unmatched to their case even though they are valid, requiring further cognitive resources in addition to seeking information~\cite{park2021}.
Exposure to such irrelevant information can delay problem-solving~\cite{cutrona1990stress} and discourage them from accessing support centers~\cite{ullman2007}.
On that account, prior findings in HCI suggest support(-seeking) systems tailor best possible support options of individual survivors~\cite{andalibi2018social}.
Through a user study with various chatbot systems, Zamora~\cite{zamora2017dave} also reports `high performing' (effective with information), `smart' (accurate and correct), and `personable' (personalizing information) as expected qualities for a chatbot.

Related to personalization, many guidelines in the practice of supporting sexual assault survivors (\eg~\cite{csom2018,koss2020,sart2011}) emphasize a \textit{victim-centered} approach through which care-givers should prioritize victim's interests, preferences, and safety~\cite{csom2018}. 
Practices of the victim-centered approach include providing tailored services to each survivor, considering their social and individual contexts given a list of available resources.
Hence, the victim-centered approach could be helpful in designing support-seeking systems for sexual violence survivors that provide personalized information.
Through exploring personalization trade-offs in designing such systems, we discuss future system implementation in light of the victim-centered approach.

\subsection{Dialogue-based Interactive Systems for Support-Seeking of Sex Crime Survivors}\label{sec:chatbot}
A useful approach to personalization systems could be dialogue-based interaction through which a system asks questions about users' needs and provides tailored information to them.
Considering the above qualities desired for support-seeking systems, recent work~\cite{park2021,ahn2020chatbot,park2020can} shows the possibility of dialogue-based interactive systems as a support channel for sex-crime survivors.
For example, Park~and~Lee~\cite{park2021} envision how sexual violence survivors find a dialogue-based interactive system's ability in lowering burdens of survivors.
In addition to saving time and cost of users, they found dialogue-based machine interaction can reduce privacy burdens due to personal disclosure and emotional concerns (\eg~mismatched cases).
They further suggest various guidelines for interface design, including camouflage interface, exporting chat log through email, etc.
Based on a focus group interview with an adult counselor and two peer advisors, Ahn~\ea~\cite{ahn2020chatbot} characterize sense-making, age-specific conversation manner and information, and openness to general public as important features for a counseling chatbot for teenager survivors.
Through analyzing questions generated by 18 people for a chatbot for reporting sexual violence cases, Maeng~and~Lee~\cite{maeng2021} suggest design implications for a hybrid reporting chatbot with rule- and machine learning-based inferences.
Zamora~\cite{zamora2017dave} indicates chatbot's absence of judgment makes users feel more comfortable to disclose.

Designers and practitioners have also utilized conversational information systems for sexual violence survivors, primarily focusing on information about reporting and articulating a case. 
For example, based in the Philippines, Gabbie~\cite{gabbie} provides brief legal information about users' cases and then leads them to a police reporting procedure. 
The information provided by Gabbie mainly includes criminal composition requirements and penalties.
Spot~\cite{spot} (in U.S.A.) prompts what to state about users' cases, which naturally informs them about what to include in their formal statements.
While these systems can help survivors to better understand their situation, survivors who are seeking support options other than juridical process (\eg~medical service, consultation) might not benefit much from them.
An Australian chat service, Hello Cass~\cite{cass} takes a relevant approach with ours.
It asks about an incident and suggests state (Victoria) or national level resource channels for particular types of victimization, yet it provides limited instructions for survivors with urgent needs.
Recently, a natural language processing-based approach has been tested for classification of sexual harassment survivors~\cite{bauer2019} in order for a system to categorize suitable resources for survivors and to support efficiently collecting data about sexual harassment.

While prior work has examined design implications~\cite{park2021,maeng2021,ahn2020chatbot,zamora2017dave} and technological implementations~\cite{calvaresi2021} for conversational personalization systems, how to design personalization of such a system has been relatively less explored.
As personalization of information is a key feature of such conversational systems, compared to other systems like a reporting or logging tool, a deeper understanding in design issues arising from personalization is necessary.
Furthermore, as the victim-centered approach and empirical studies in HCI~\cite{andalibi2018social,park2021} imply, sexual violence survivors are people with different characteristics rather than a single user group.
In other words, ``the best'' solution may not exist, and designers and practitioners might need to understand key design trade-offs and seek balance around them.
However, little is known around how conversational information systems for sexual violence survivors should understand users for personalization and inform them about such tailored information.
To facilitate more of such systems, therefore, we explore design trade-offs arising from personalization by learning through design workshops with expert care-givers (counseling and medical experts) and an interview study with targeted users (sexual violence survivors).

\section{Method Overview}
We aim at exploring personalization trade-offs by designing a dialogue-based system that provides personalized support-seeking information for sexual violence survivors. 
As designing for stigmatized people requires both profound domain knowledge and perspectives of end users, we had multiple stages of participation (\autoref{tab:method}): 
(1) sensitization with various sources; (2) four design workshop sessions with three expert care-givers; (3) focus group interviews with four domain experts; and (4) interview study with four self-identified sexual violence survivors.
Our methods were approved by the IRB of the researchers' institution.
We used Korean for our studies, so direct quotes are translated to English.

\begin{table*}[t]
    \centering
    \includegraphics[width=\textwidth]{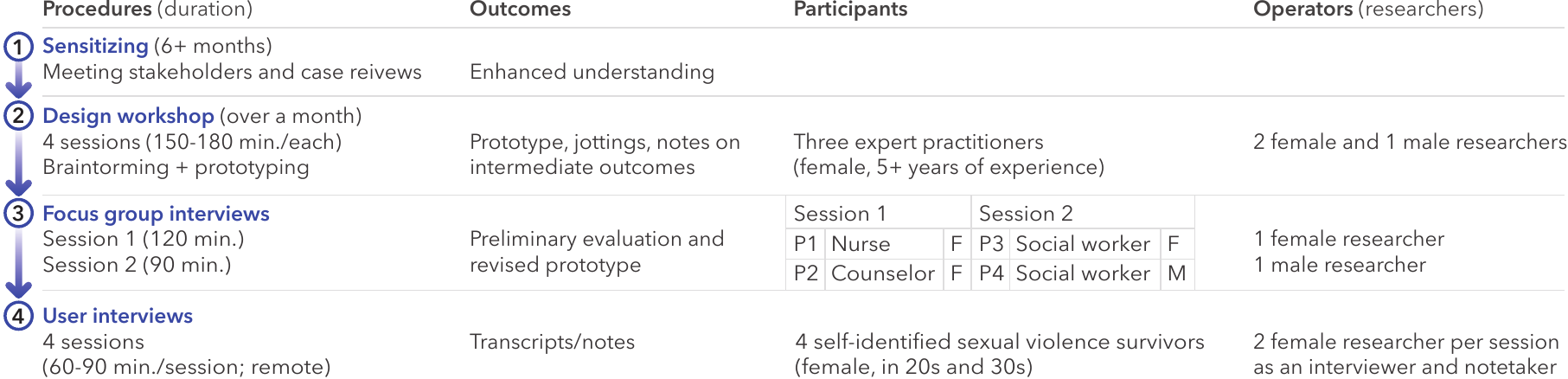}
    \caption{Overview of the entire process of our study. We provide aggregate information for interview participants to prevent potential re-identification. F(emale), M(ale).}
    \Description{This table has four columns and four body rows. The columns are Procedures (duration), Outcomes, Participants, and Operators (researchers). This description reads each row in the column order. In the first row, procedures (duration) are: Sensitizing, over six months, meeting stakeholders and case reviews. Outcomes are: Enhanced understanding.
    Participants and Operators are empty for this row. In the second row, Procedures (duration) are: Design workshop, over a month, four sessions, 150 to 180 minutes each, Brainstorming and prototyping. Outcomes are: Prototype, jottings, notes on intermediate outcomes. Participants are: Three expert practitioners, female, more than five years of experience. Operators (researchers) are two female and one male researchers. In the third row, Procedures (duration) are: Focus group interview, Session 1, 120 minutes, Session 2, 90 minutes. Participants are: In Session one, P1, Nurse, Female and P2, Counselor, Female. In Session two, P3, Social Worker, Female, and P4, Social Worker, Male. Operators are: one female researcher and one male researcher. In the fourth row, Procedures (duration) are: User interviews, four sessions, 60 to 90 minutes per session, remote. Outcomes are: Transcripts and notes. Participants are: Four self-identified sexual violence survivors, female in 20s and 30s. Operators (researchers) are: tow female researcher per session as an interviewer and note-taker.}
    \label{tab:method}
\end{table*}

\bstart{Sensitizing}\label{sec:sensitization}
We initially set our scope as informational assistance for sexual violence survivors through iterative informal discussions with field officers and policy makers in police, community service, and justice departments of the Republic of Korea, which was reassured by the workshop participants.
Then, we sensitized ourselves to the topic before actual designing in order to prevent ourselves from making mistakes due to poor awareness.
Sensitization helped us to be on the same page without forcing any of us to disclose their own experiences.
We spent more than six months reviewing legal cases and news reports; meeting survivor-activists; meeting counselors and police officers; and making several low fidelity prototypes. 
To reduce our bias as much as possible, we triangulated our sensitization with previous studies on support-seeking of sexual violence survivors, as provided in \sectionref{sec:intro} and \sectionref{sec:background}.

Sexual violence survivors and activists we met during the sensitization informed us that many survivors are tired of being requested to participate in studies because they have often had difficulty in identifying agents who are conducting a study.
In addition, they were seldom able to see a substantial product or report as an outcome of the study they have participated in, or they have found such an outcome incongruous with their intentions. 
Thus, we shaped our study to begin with collaboration with experts and then extend it to end users.
Our collaborative design workshop with experts was effective in building rapport and trust with them (and their institution), which helped us to reach out survivors for interview by ensuring the security of their participation.

\bstart{Design workshop, focus group interview, and user interview}
After sensitizing, we had four design workshop sessions with three expert care-givers (\autoref{sec:workshop_participants} and \autoref{sec:sessions}).
Through this workshop, we derived high-level design considerations for our prototype system (\autoref{sec:considerations}) and then implemented a prototype conversational system for support seeking of sexual violence survivors (\autoref{sec:implementation}).
Using our prototype, then, we conducted two focus group interviews with four domain experts (two per session) to reduce potential harms of our prototype against target users as much as possible before we meet them (\autoref{sec:preliminary_evaluation}).
To expand our understanding from the earlier steps and evaluate our prototype with target users, we conducted semi-structured interviews with four self-identified sexual violence survivors (\autoref{sec:userstudy}). 
Further details are described in the referenced sections.

\bstart{Position statement}
While we do not claim preferability of a specific interaction format, we focus on a dialogue-based interactive system for sexual violence survivors as a test bed for its popularity and potentials.
In the region of expected users (Republic of Korea), 87\% of people (about 45 million out of 52 million~\cite{kakaousers}) use the same messenger application that provides various information chat services at different levels of community (\eg~local township, national services, \etc).
For such high popularity of the format, stakeholders whom we met during our sensitizing process (\autoref{sec:sensitization}), co-design workshop (\autoref{sec:workshop}), and user studies (\autoref{sec:userstudy}) recommended a dialogue-based system.
Along with the reasons above, prior studies in various domains (\eg~\cite{park2021,Heerden2017,wang2018}) have adopted a dialogue-based approach to ensure qualities like anonymity, wide accessibility, and promptness that are also desired in our case (\autoref{sec:support}).

Next, we intend our system to provide information about ``access'' to available professional care-givers (\eg~experienced social workers, therapists) tailored for individual cases, rather than an intervention system that can potentially replace those care-givers.
The scope of our system includes contact information to care-givers, before-visit instructions (\eg~not drinking liquid), high-level guidance of using support resources. 
Our prototype does not have features for actual intervention (\eg~diagnosis, reporting, recording, consultation, legal advice) by any means.

Legal and institutional information in this paper is based on that of the Republic of Korea.
In general, sexual violence is primarily treated as a criminal case, and survivors can file a request for compensation as part of the criminal case under the related law.
Workplace verbal harassment is treated as an administrative case.
Rape myth is quite pervasive in Korea, so offenders often strategically countersue survivors for false allegation just to burden those survivors and prolong the case~\cite{jinsook2018,minsook2017}.
Core support resources for survivors are usually operated or funded by the government, yet there are non-government organizations that independently provide relevant support resources.

\section{Design Workshops with Experts}\label{sec:workshop}
The primary goal of our design workshop was to understand design concerns from operation of support services through actually implementing a prototype with professional caregivers.
We collaborated with domain experts in integrated support (medical care, walk-in process, counseling) for sexual violence survivors through four sessions of design workshops.
We then conducted preliminary evaluation with four experts to ensure the overall usability and correctness of our prototype.
As an outcome of our design workshop, we illustrate our prototype with design rationales.
In this section, `we' denote both researchers and workshop participants as a collaborative design team.

\subsection{Method}

\subsubsection{Workshop participants}\label{sec:workshop_participants}
\bstartnc{Our expert participants}, Megan, Vivian, and Jennifer (pseu\-do\-nyms; female), have been affiliated with the same sexual assault support institution in the Republic of Korea for more than five years.
Their institution is a branch of a nation-wide organization that provides integrated support for sexual violence survivors. The institution follows a specific sequence of procedures as a person walks into this institution: emergency support, medical assistance, and long-term consultation. On this account, we recruited three professionals, one from each department, to enrich our discussion. Megan provides emergency counseling and determines the next steps for visitors based on their cases. Her team is the first contact point of all the visitors. Vivian is an inbound nurse who inspects the visitors' medical status, collects physiological evidence, and plans further medical support if needed. Finally, Jennifer works as a long-term counselor and helps survivors to overcome their PTSD symptoms. Megan and Vivian were primary information sources because they are the first contact point and a medical expert, respectively. While we decided not to aim for longitudinal consultation during the workshop, Jennifer informed us about how the earlier acquisition of information can affect the survivors' later actions.
Each of them was compensated approximately USD 180 per session according to the compensation guidance for professional consultation of the researchers' institute.

\bstartnc{Researchers} who participated in the workshop were two female and one male researchers (expertise in HCI; in their 20s). With perspectives from user interaction and design, they provided possible options to realize (but not limit) what experts discussed. For example, when experts mentioned needs for age of a user, then the researchers suggested possible options for how to ask age (\eg~actual age or age group).

\subsubsection{Workshop sessions}\label{sec:sessions}
Inspired by prior work on design research~\cite{pederson2015war,peffers2006design}, the researchers set the following agendas: (1) finding requirements, (2) detailing requirements, (3) designing information, and (4) building questions.
We iteratively adjusted plans after each session and allowed for flexibility during sessions.
For example, we initially planned building questions before designing information, but we swapped those session themes after realizing that dialogue questions are contingent on system-provided information.
We discuss more about this in \autoref{sec:reversed_approach}.
We were not able to voice-record the workshops because recording the participants' cases may violate the participants code of conducts as they unavoidably had to discuss based on their previous cases.
Instead, we kept notes in each workshop session about the workshop activities and their outcomes.
Each session took approximately 150-180 minutes. 

\bstart{Session 1--finding initial requirements}\label{sec:session1}
The primary tasks in Session 1 were to define a goal and find initial requirements.
We started with sharing high-level domain knowledge of the participants (support process for survivors) and researchers (designing an interactive system).
After setting our initial goal as assisting ``access'' to professional support, we had an affinity diagram activity for requirement finding, where we had no guidelines except for the overall topic: support for survivors.
Based on the affinity diagram, we derived possible phases (a chunk of conversation with the same functionality) of a dialogue system for support-seeking of survivors.

\bstart{Session 2--detailing requirements}\label{sec:session2}
In Session 2, we focused on detailing requirements derived in Session 1 by discussing what to add or remove and how to order phases of dialogue.
For example, we removed a part about informing how to `write' a formal statement (or affidavit) because misunderstanding it may exacerbate their situation; instead, we included a section about telling how a formal statement `functions' in the course of problem solving.
In addition, we prototyped a greetings and introduction phase using the low-fidelity prototyping tool (\sectref{sec:prototyping_tool}), which provided the participants with a clearer image of dialogue-based interaction.

\bstart{Session 3--designing information}\label{sec:session3}
Upon reviewing the results from Session 2, we needed to scope a set of information to provide before generating questions for tailoring.
We were concerned that if we set questions first, then those questions might limit the range of information to be provided later, or we might include unnecessary questions that demands excessive disclosure. 
Thus in Session 3, we brainstormed what information to provide using sticky notes and then discussed whether, how in detail, and upon which condition to provide them.

\bstart{Session 4--question building}\label{sec:session4}
In the final session, we specified questions to include so that the system can tailor information based on the conditions derived in Session 3.
The conditions and questions were not one-to-one matched because some conditions consisted of sub-conditions and conflicted each other.
For instance, when a perpetrator is younger than 13 years of age, then many other conditions become unnecessary because a child under 13 is exempted from criminal liability. 
After creating a list of questions, we prototyped them using the low-fidelity prototyping tool.

\begin{figure}[t]
 \centering
 \includegraphics[width=\columnwidth]{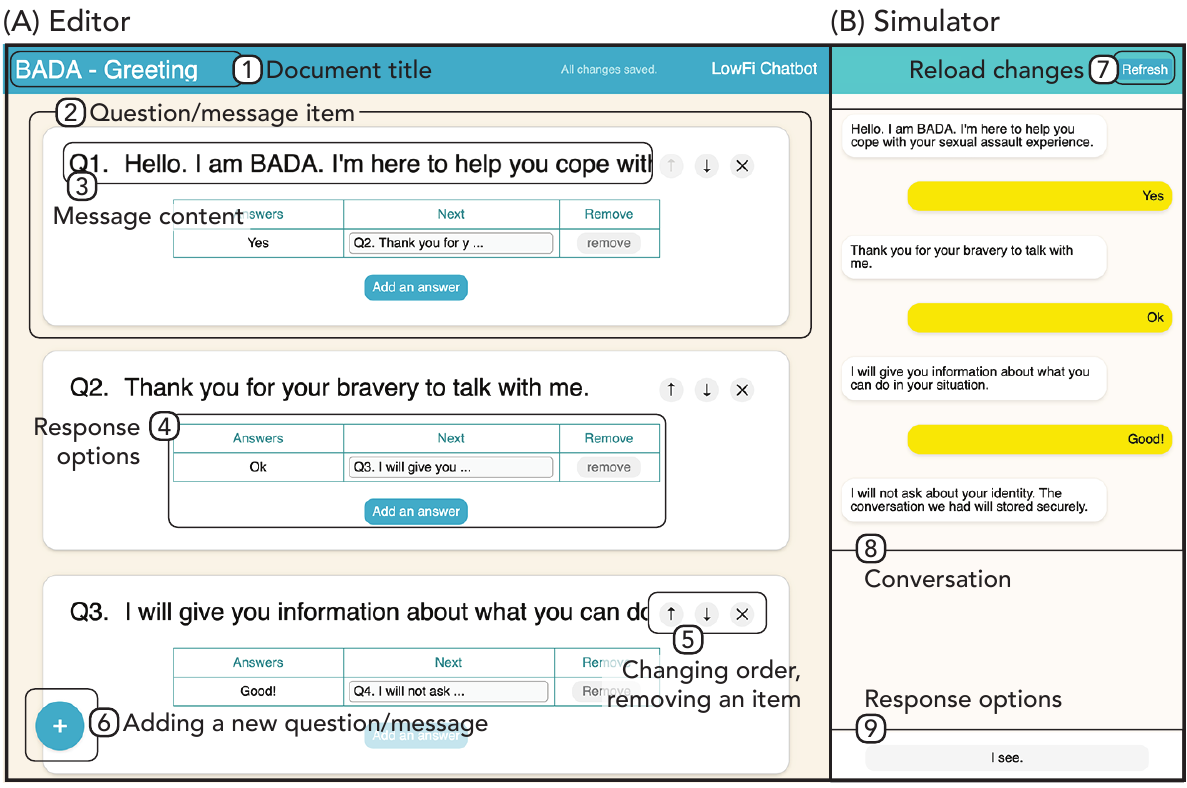}
 \caption{A screenshot of our low-fidelity prototyping tool.}
 \Description{this figure contains an application interface of our low-fidelity prototyping tool described in Section 4.1.3. The interface has two sections. On the left, Section A is the editor, and on the right, Section B is the simulator. In Section A, the top row contains the document title, notated by (1). Below, there are lists of three demo question/message items. These items have the same structure, but the number notations are spread across the items for convenience. In the first question/message item of the list, noted by (2), there is title, saying "Hello, I am Bada. I'm here to help you cope with ...", notated by (3). In the second item, there is a table of response options, notated by (4), which has columns of "Answers", "Next", and "Remove". Each row has answer expressions, next item to go, and remove buttons. In the third item, there are down arrow, up arrow and cross buttons on the top right corner of the item, notated by (5). They are for changing the order of the item upward or downward, and removing the item. On the left bottom corner of Section A, there is a circled plus button, notated by (6), for adding a new question/message item. In the Section B, the simulator is illustrated. The top row contains the refresh button, notated by (7). Below, a sample conversation, notated by (8), is illustrated. At the bottom, there are response options at the moment, notated by (90.}
 \label{fig:lowfi}
\end{figure}

\subsubsection{Low-fidelity prototyping tool}\label{sec:prototyping_tool}
In addition to ideation activities using sticky notes, we developed and utilized a low-fidelity prototyping tool depicted in Figure~\ref{fig:lowfi}. The left section (A) of the prototyping tool is an editor of a dialogue storyboard, and the right section (B) is a simulator. The usage of the editor is quite similar to that of Google Forms. A user can set the title of a storyboard (1). Then, the user adds a question (6 $\rightarrow$ 2, 3) and attaches available answers (4). Each answer option can be directed to another question. The order of questions can be modified by up/down buttons (5). After or while editing, the user can run through their storyboard (8 \& 9) by refreshing the simulator (7).

We used this tool to provide an intuition of a dialogue-based interaction to our participants rather than overwhelming them with complicated diagrams. 
Although interactive conversations are often not linear, the linear representation of the storyboard and instant simulations seemed to help the participants understand how a chatbot works. 
It also encouraged the participants to project their ideas envisioning our prototype. 
For example, when we met for Session 3, the three participants brought a list of questions that they had made voluntarily. 
This happened after we used the prototyping tool in Session 2. 
To prevent from limiting the participants' ideas within this tool, we iteratively employed this tool and sticky notes for low-fidelity prototyping and brainstorming, respectively.

\subsubsection{Preliminary evaluation}\label{sec:preliminary_evaluation}
We performed two focus group interviews (in-person) with four domain experts (P1--4, average age of 36.8) for preliminary evaluation.
As shown in \autoref{tab:method}, P1 (nurse, female) and P2 (counselor, female) from the same institution for sexual violence survivors participated in Session 1. P3 (social worker, female) and P4 (social worker, male) who had worked with sexual violence survivors joined Session 2.
Session 1 (conducted by a female researcher) took about two hours, and Session 2 (conducted by a male researcher) took 90 minutes.
In each session, we asked to use our prototype for 10 minutes on their smartphones and then asked several questions regarding usability, appropriateness, correctness, and further suggestions. 
In general, the expert interviewees positively evaluated our system in terms of appropriateness and correctness. 
Each of them was compensated approximately USD 85 according to the compensation guidance for professional consultation of the researchers' institute.
We describe reflections their suggestions for improvement in \autoref{sec:implementation}.

\subsection{Design Considerations}\label{sec:considerations}
Through the design workshops, we derived the following design considerations to motivate the implementation of our prototype conversational system.

\subsubsection{High-level system considerations}
We discussed rationales for high-level design of our prototype conversational system regarding a preferred state, conversation length, user group, and speech tone.

\cons{Help users to initiate actions with envisioning consequences}{c:initiate}
We framed the preferred state after using our system as `to be ready to access professional support that they need or want to visit,' instead of vague framing like `to provide relevant information.'
We prioritized informational readiness rather than emotional assistance because the workshop participants doubted whether emotional assistance is a \textit{right} thing, considering the needs for short initial contacts (to promote next steps) and side effects of emotional mismatch, such as refusal of seeking support~\cite{davis1991supportive,siegel1990reactions}.
The workshop participants indicated that uncertainty about next steps often delay survivors from visiting them.
Thus, our system aims at helping survivors to visit professional support while being able to envision the outcomes of their actions.

\cons{Be prompt}{c:prompt}
The workshop participants all emphasized the needs for short dialogue to encourage immediate actions and aftercare.
When the researchers suggested 20 minutes of time window for stable use, Megan disagreed because her team tries to complete the initial contact in five minutes to provide timely support, and Jennifer seconded the importance of promptness. 
For example, immediate action is highly required to obtain ephemeral, critical evidence, such as CCTV records and medical test results, and to enhance effects of medical treatment.

\cons{Be prepared for unexpected users}{c:unexpected}
We set our primary user group as adult sexual violence survivors (and older teenagers) because children survivors' misunderstanding can be more detrimental than that of adult survivors (\eg~their statement can be declined as evidence).
However, if we deployed this system for public, underage users would still be able to access our system.
Hence, our system attempts to guide those unexpected users such as underage survivors.

\cons{Have a credible speech tone}{c:credible}
A convincing manner is demanded to help survivors feel ready to access professional support.
Jennifer and Megan said that they use a polite tone with adult survivors to display a credible impression to better ensure that survivors keep in touch with them.
Prior findings in computer-mediated communication~\cite{morand2003politeness,lam2011linguistic} and conversational agents~\cite{folstad2018makes} also suggest that a polite, professional tone of a system may help to build trust with users.
Thus, our system uses a polite, credible tone for our target user group (adult survivors).

\subsubsection{Considerations for users' understanding}
Sexual violence survivors' clear understanding of situation and related information is important to effectively solve their problems and fulfill their needs.
We identified misinformation and vagueness as challenges against clear understanding of users for our system; Yet, we do not claim that these are the exhaustive set of such challenges.

\cons{Be correct according to users' cases}{c:correct}
The workshop participants stressed correctness is important because pervasive misinformation can severely disturb survivors.
For example, Vivian mentioned that survivors often ask for diagnosis of sexually transmitted infections to collect physiological evidence, but they become disappointed after she informs that the diagnostic test cannot be used as evidence.
Correctness of information depends on individual cases.
Megan said that her initial contact team categorize their clients because available solutions differ by cases, and Vivian, a nurse, added that such categorization helps her to optimize medical solution.
Thus, our system personalizes information based on users' response about their cases.

\cons{Be direct}{c:direct}
Providing correct and appropriate information is viable through a clear conversation.
The workshop participants, for example, have used yes/no questions and direct expressions to avoid their and clients' misunderstanding of a situation in practice, which was also supported in our preliminary evaluation.
They indicated that yes/no questions can help with making a prompt progress and finding suitable support methods.
In addition, they noted that they try to use direct terms like `rape' during their in-person counseling because euphemisms like a `bad thing' are likely to delay a consultation process.
Prior work reports that direct wording is likely to increase the validity of data collected from sex crime survivors by minimizing their misunderstanding of investigation~\cite{krebs2011comparing}.
We were also concerned with misunderstanding when a client and counsellor use the same term with different meanings (as they are non-daily words).
Whereas counselors can adjust their tones during in-person conversations, systems aimed for public deployment like ours tend to have a limited control on users' interpretation during a one-time, short conversation.
Thus, our system uses direct expressions to clearly ask questions and deliver tailored information to users.

\subsubsection{External constraints}
There are legal and practical constraints that we had to address in our system regarding data storage and temporal availability of resources.
We do not claim those constraints are only external constraints. They may vary in different regions and can change in the future.

\cons{Store data but be anonymous}{c:storage}
Securely storing the original data is legally required for record purposes in case where a conversation needs to be verified (\eg~providing it as a proof) and is a common practice at counseling providers, including our workshop participants'.
This never implies that actual counselors do not trust their clients.
However, Jennifer indicated storing private data may discourage survivors, so we decided not to collect direct identification information (\eg~email, date of birth), which can limit such verification procedures.
Prior findings~\cite{park2021,naseem2020designing} also point out that data storage can be harmful to survivors when abusers have access to their devices or usage logs.
Thus, our system has an anonymous reference protocol for future verification.

\cons{Consider temporal availability of resources}{c:temporal}
In practice, support resources are provided for those who are in particular time after incidents.
For example, if a user specifically indicates that they wants to file a case, tailoring information must be in line with the statute of limitations (the period of time only during which a crime suspect can be indicted) because after its termination, the number of available legal options decreases significantly. 
Similarly, medical support options are mostly provided to those in urgent cases (\eg~a survivor cannot ask treatment for what happened several years ago). 
Thus, our system considers temporal availability of those resources when personalizing them.

\subsection{Prototype Design}\label{sec:implementation}
As an outcome of the design workshop, we implemented a prototype conversational system for support-seeking of sexual violence survivors.
While we did not deploy our system, we aimed at a production version to observe realistic design problems (\eg~data storage).
The workshop participants wrote most of the questions of our prototype, and the researchers drafted expressions for provided information, which the workshop and preliminary evaluation participants eventually validated.
We provide a video demo in \url{https://osf.io/qyscg/}.

\subsubsection{Goal and scope}
The goal of our system is `to provide personalized information about possible actions the survivors can take and their possible outcomes,' such as how to preserve remaining evidences, advantages of making consistent notes, and an outline of reporting procedures (\ci{c:initiate}: help initiation with envisioning).
To motivate users to access professional support channels, we added phone numbers, addresses, and web sites of relevant support institutions that are available in the location they responded (\ci{c:correct}: be correct with users' cases).

Although adult survivors are our primary target users, we added dialogue for underage users (\ci{c:unexpected}: be prepared for unexpected users).
For instance, when users identify themselves as underage, we included a message that encourages to use the system with trusted adults, such as parents or teachers, as well as contact information about specialized institutions for children survivors.

During the workshop, we tried to focus on one-time sexual assaults, including rape (with drug or alcohol), workplace sexual harassment, revenge porn, and (having a history of) child sexual abuse, which are generally consulted at support institutions according to our participants.
We did not directly include long-term, repeated assaults, such as longitudinal domestic violence and sexual grooming because repeated assaults often need further judgment of prosecutors and judges; yet, they are comprised of and hence detectable by short-term assaults as it is further investigated.

\subsubsection{Interface design}
\begin{figure}[t]
 \centering
 \includegraphics[width=\columnwidth]{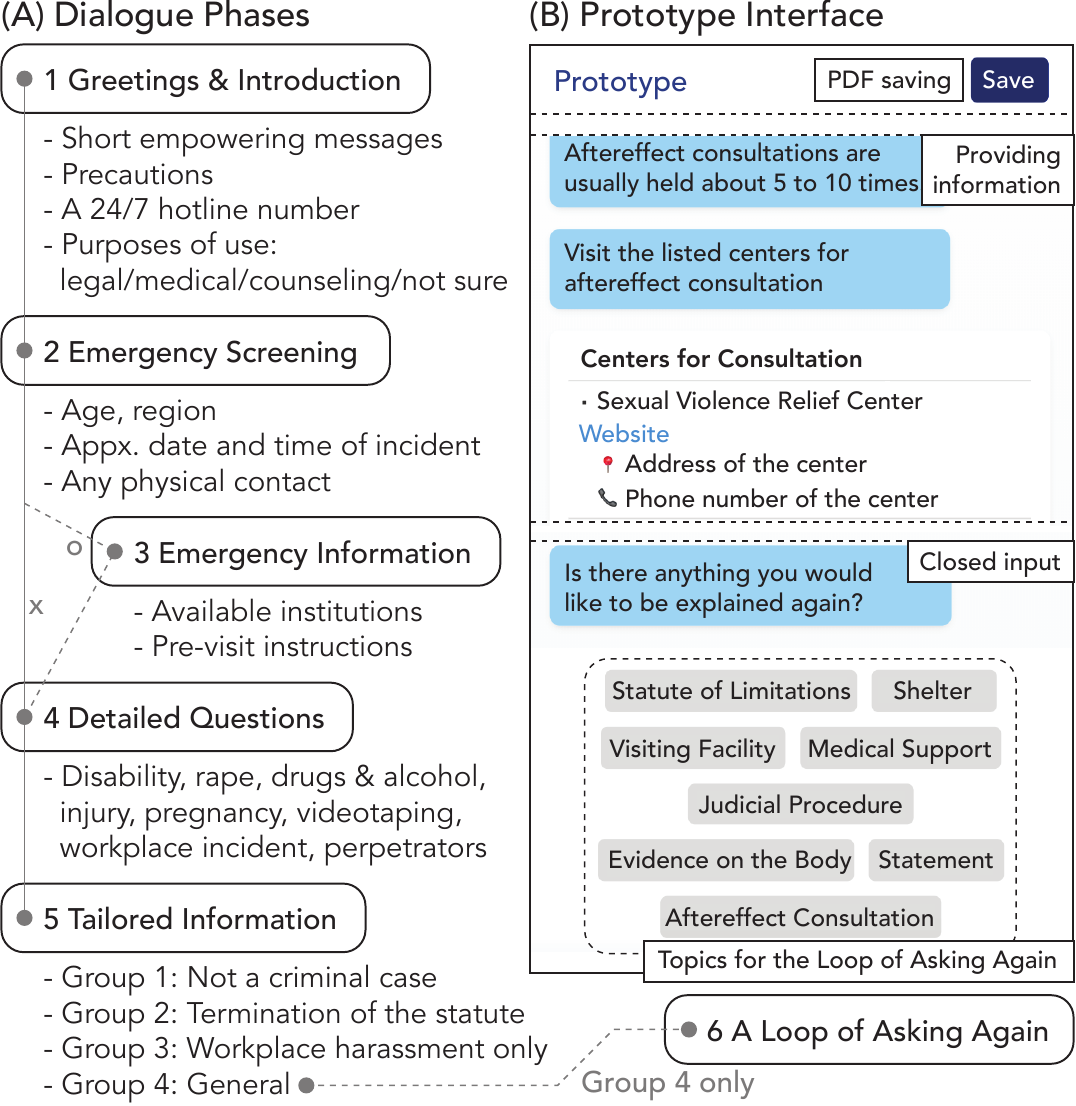}
 \caption{Prototype implementation. (A) An overview of dialogue phases. (B) The user interface of out prototype.}
 \Description{This figure has two columns, A and B. In the column A, a diagram describes the conversation procedure of our prototype chatbot. The diagram has six phases: (1) Greetings and introduction, (2) Emergency screening, (3) Emergency information, (4) Detailed questions, (5) Tailored information, and (6) A loop of asking again. These phases are further described in the Section 4.3.2. The right column (b) shows the prototype interface design, featuring the save button, a sample conversation, and user response options for Phase 6, a loop of asking again.}
 \label{fig:prototype}
\end{figure}

Our prototype interface is illustrated in \autoref{fig:prototype} (B).
To overview, a conversation is expected to be done within five to ten minutes in a close-ended format and direct speech tone.
Our prototype has six phases of conversation: (1) Greetings and Introduction, (2) Emergency Screening, (3) Emergency Information, (4) Detail Questions, (5) Tailored Information, and (6) Loop of Asking Again as outlined in \autoref{fig:prototype} (A).
Users can download a PDF file of their conversation with the system.

\bstart{Time window}
We intended the time window (an expected time duration for a single use) to be five to ten minutes to encourage immediate actions and aftercare (\ci{c:prompt}: be prompt).

\bstart{Speech tone}
Our prototype uses direct expressions and maintains polite speech tone to ensure clear understanding and trustworthiness (\ci{c:direct}: be direct; \ci{c:credible}: be credible).
To be direct, we tried to formulate questions with clear expressions and avoided words with non-daily, legal meanings.
For example, our system expresses `rape' as penetration to the user/perpetrator's sexual organ.
To have a trustworthy tone, we used speech forms that are widely used in a formal situation and between those who meet for the first time in Korean language (the `Seubnida' and `Haeyo' forms).

\bstart{Close-ended format} 
The needs for short time window and tailored information motivate a close-ended format to enhance easier response and prevent the system from misinterpreting user responses (\ci{c:prompt}: be prompt; \ci{c:direct}: be direct).
This decision led us to using a rule-based approach for personalization rather than a machine learning-based approach.

\bstart{Data download} 
Storing the chat log is useful for survivors to review the log in the future~\cite{park2021} and legally required in case of future verification (\ci{c:storage}: store data anonymously).
To address this issue, we decided not to ask and store any direct identification information (\eg~email) that can identify someone by itself only rather than having a powerful authentication method.
Instead, we use `timestamp' as a verification reference, which can potentially be scaled up to a timestamp-based hash string.
As such anonymous storage can limit users from reviewing their conversation, our prototype lets users instantly download their dialogue in a PDF format at any time while using it, and its short time window makes it relatively easier to obtain the same information later.

\bstartnc{In Greetings and Introduction phase}, our prototype gives brief empowering messages, precautions, and a hotline number, and then it asks the purpose of use.
The precautions involve the storage of conversation, the unavailability of reaccess, how to download their chat logs, and the expected time window.
Next, it gives a 24/7 hotline number for those who do not want to go through the entire chat to further shorten the time window (\ci{c:prompt}: be prompt).
Finally, the user can select one or more purposes of use among reporting/investigation, medical support, aftereffect consultation, and ``I am not sure'' for those who have not clearly specified their purposes.
To justify explicit expressions, our prototype says it `might remind you of unpleasant memories regarding the incident ... to guide you correctly.'

\bstartnc{In Emergency Screening phase}, our prototype asks questions about two screening conditions for emergency: (1) happened within 5 days (2) with physical contact (\ci{c:prompt}: be prompt).
We included this phase to quickly inform users about critical actions that they need to take at the moment (\ie~visiting any support channel is much more important than keep using our system).
In particular, after asking the user's age and region (for filtering the available institutions), our system inquires the approximate time of occurrence, and any physical contact (for screening). 

\bstartnc{Emergency Information phase} is activated if a user is screened as under emergency.
This phase gives a list of available institutions for medical and legal support (\ci{c:correct}: correctly personalize), followed by pre-visit instructions to help users better envision their next steps (\ci{c:initiate}: help initiate).
For example, according to guidelines from our workshop participants' institution, when a case happened within 72 hours, they are informed about an alcohol and drug test.
For cases within 12 hours ago, the user is recommended to drink or eat nothing to preserve physiological evidence remaining in their mouth.
These are crucial for survivors in urgent cases to receive timely medical and legal support. 
Finally, our prototype encourages users to keep using it for more information on their way to a suggested institution.

\bstartnc{In Detailed Questions phase}, our prototype asks whether the user has or experienced any of the following: disability, rape, the use of drugs and alcohol, physical injury, pregnancy, digital violence, workplace harassment, and perpetrators, in this order.
We chose those question items by first categorizing support resources and then deriving conditions to personalize those support resources (see \autoref{sec:reversed_approach} for details).
For example, questions about the use of drugs and alcohol and physical injury are used to prioritize relevant medical instructions and resources. 
Questions about disability, digital violence, and workplace harassment are used to suggest more suitable support providers (\eg~specialized counseling, removing nonconsensual videos).
This process helped us to not include unnecessary private questions like `what weapons did the perpetrator use?' that we had in our sensitization period.
While we did not include an explicit option for skipping, they could answer ``I am not sure'' to most of those questions.
Our prototype activates this phase when it is certain that the statute of limitations (10 years) is not reached (\eg~more than 15 years) because support resources personalized by such information are unavailable after passing the statute of limitations (\ci{c:temporal}: temporal availability).
We revised many expressions after preliminary evaluation to remove questions that are unnecessarily private and can potentially stigmatize users. 
For instance, we modified ``where do you live?'' to  ``where do you think you can get support better?'' so that collected data do not indicate users' residence (P2, P3); ``is drug suspected?'' to ``do you think the perpetrator used drugs?'' to prevent self-blaming of users (P4).

\bstart{Tailored Information phase}
After the user responds to the questions in Phases 2 and 4, such as user's case details, region (not residence), age, and, the time of occurrence, our prototype filters in and provide relevant information (\ci{c:correct}: provide correct information with personalization).
It first classifies the user into four groups: (1) not making a case, (2) the termination of the statute of limitations, (3) workplace (verbal) harassment only, and (4) general. 
Our prototype divides those four groups based on availability of support options for them.

Since most medical, legal, and urgent services are not intended for Group 1, 2, and 3 users in the Republic of Korea, our prototype describes their situation and suggests a few available resources for them.
Users are classified to Group 1 (not making a case) when they have answered explicitly ``no'' (different from ``I am not sure'') to every question about case details.
For Group 1, the system recommends nearby PTSD consultation institutions in case when they were not able to respond successfully for any reason.
When the termination of the statute can be clearly determined (for more than 15 years after the incident unless it had happened when the user was an underage), the system classifies the user to Group 2 (\ci{c:temporal}: temporal availability). 
In addition to nearby consultation institutions, we suggest Group 2 users to visit a nearby police department for more accurate calculation of the statute of limitations.
If a user indicates only verbal abuse at a workplace, the system categorizes them into Group 3, and recommends a consultation institutions specialized for workplace sexual harassment.

For Group 4, our prototype first provides contacts to available shelters, support institutions, and regional police departments based on their regions with pre-visit instructions to help users actually contact them.
Our prototype informs about overall procedures of such providers to help users envision about them (\ci{c:initiate}: help initiate and envision).
Then, it describes the statute of limitations and how to know it (\ie~visiting a nearby police department).
Our prototype does not directly calculates the statute of limitations to prevent false expectation because it requires professional interpretation (which may require a legal judgment).
Next, our prototype provides information about types of physiological evidence and how to collect it, primary information about writing statement, available medical services they can opt in, aftereffect consultation, and an overview of legal process, in this order.
We ordered this in terms of temporal priority (\ie~what they have to do earlier), assuming that they are not in urgent cases which are screened and informed in earlier phases.
To enhance the relevancy of information, our prototype flexibly adjusts the order of information.
For example, those who indicate medical support (but not legal options) as their purpose, then our prototype first provides medical service information prior to statement and evidence collection.
Our system illustrates how aftereffect consultation and legal process will be like if they plan to take relevant actions (\ci{c:initiate}: help envision).
We also revised expressions after preliminary evaluation, such as clarifying the meaning of ``tracking inspection'' as a gynecologist inspection method (P4), and removing information about support for in-hospital treatment which is often less likely (P1).

\bstartnc{In Loop of Asking Again phase}, Group 4 users can ask repeatedly about their interested topics, as the amount of information provided to Group 4 (about 7-8 short paragraphs) is relatively longer than that to the others (skimmable length).
We added this phase after the preliminary evaluation with experts, suggested by P4, in case where users develop different priority or interests after seeing the provided information.
When a user chooses an irrelevant topic, our prototype informs that it seems less relevant to them. 
For example, if a user who experienced rape two years ago requests information on ``evidence on the body'' topic, our system first proposes that ``In the current situation, it is difficult to collect evidence that remains on your body. 
Select item `Statement' to see more related information. 
But for your information, we give you the information below.''

\subsubsection{Technical details}
We built our prototype on a Node.js environment using Express.js and Vue.js as back-end and front-end frameworks, respectively. 
Our prototype manages conversation using explicit rules (\ie~no black box) because conditions for asking questions are logically expressible in our case, and because we avoid probabilistic conditioning of information to prevent potential mismatch and false expectation.
As a user proceed, our prototype updates a feature dictionary to support processing those rules.
Some features are straightforward which only needs a single question (\eg~whether it was 12 hours ago, who was the offender), and other features are composed of multiple subordinate features (\eg~rape by family member, emergency).
Each sentence is provided when its logical conditions are saturated by the feature dictionary (\eg~asking consents about video recordings when there was a video recording).

\section{User Interviews}\label{sec:userstudy}
Using our prototype dialogue system, we conducted semi-structured interviews with four self-identified sexual violence survivors to deepen our initial understanding with target users as well as user evaluation.

\subsection{Method}
\bstart{Participants} We recruited participants by snowballing through caregivers, which took five months due to the COVID-19 pandemic.
In other words, they are those who have sought professional support.
Interview participants were all female and in their 20's or 30's, and are denoted as U1 to U4\footnote{We provide only aggregate information of our participants to reduce any chance of reidentification.}.

\bstart{Procedures} We used Zoom due to the pandemic with the participants' video muted, yet we did not mute researchers' video to enhance trustworthiness.
In each session, a researcher operated the interview, and another researcher took notes of the participant's response.
The interviewers and note-takers were all female researchers (expertise in HCI, in their 20s).
After consenting, the interviewees used our prototype for about 10 minutes with a provided link on their device.
We then asked ``What is your overall impression of prototype?'', followed by questions about sensitivity and relevancy like ``When you were using the prototype, what have you found sensitive or uncomfortable?'' or ``How relevant was the information provided in the prototype based on your response?''; ``Was the conversation convincing or trustworthy?''; and ``How would you improve our system?''
Each interview took 60 to 90 minutes, and they were reimbursed approximately USD 165, which was a reasonable financial amount and suggested by the caregivers we consulted for the recruitment.

\bstart{Interview protocols}
To best ensure privacy and safety of the participants (\eg~information leak, possible traumatic experience), we set up the following protocols from recruitment to interview and reimbursement. Before interview and while recruiting, we asked caregivers to simply provide overview of our study and our contact information to interested survivors and not to imply or suggest that they should participate in order to ensure voluntary participation.
We provided an option of inviting a third-party chaperone or therapist during the session (no one requested).

During each interview session, we informed the interviewee about withdrawal of participation without telling a reason (no one withdrew) and about possible aftercare options (\eg~counseling, monetary support) in case of any traumatic experience during interview (no one requested).
We recorded the interview only when the interviewee explicitly agreed with recording (one session was not recorded, so we took notes instead).
We did not record their usage of our prototype.
We did not request to use our prototype based on their past experience nor ask them to reveal any of their past experience, yet some of the participants voluntarily responded based on their past support-seeking experience (not from their cases).
We never corrected their opinion or interpretation of our system.

After an interview session, we did not inform the caregivers who helped recruitment about whether the introduced survivors actually participated while we appreciated the caregivers.
We did not provide personal information of participants and the study context to the finance department for the reimbursement.
We provided a link to an anonymous survey after completing every interview session to help them express what they could not say (no one responded).

\bstart{Analysis}
Using thematic analysis \cite{braun2006thematic}, we analyzed transcripts and notes from the interviews in both deductive and inductive ways.
We initially had questions related to major interests arising from trade-offs we characterized during the design workshop (sensitivity vs. specificity and relevancy vs. inclusiveness), so we did a deductive analysis to address those interests.
In particular, after each interview, the researchers first had online and offline discussion about the session, iteratively deriving high-level insights in light of our interests from the design workshop, and then we started an iterative coding process with those insights in mind.
Sensitivity, relevancy, and actionable instructions are such deductive themes.
However, we did not limit our themes to our initial interests. 
As we iterate our coding process, new themes emerged from our interviewees that we did not expect beforehand, such as accountability and difficulty.
\subsection{Interview Findings}
Our interviewees in general agreed that our prototype provides correct and clear information in a polite and convincing manner.
For example, U4 mentioned that a widely available system that is similarly designed to ours could help survivors to reduce time to access support channels, compared to what she had had to do without a proper information channel (\ie~asking several people about support resources).
Below, we describe how our interviewees reacted to our prototype in terms of existing approaches, comfortability, relevancy, difficulty, and actionable instructions.

\subsubsection{Existing approaches}
While our interviewees had sought professional support primarily through trusted others, they (U1, U3, and U4) had had to take multiple steps before actually getting in touch with a professional support channel.
For example, U1 said that she had become to know about a counseling institution after her acquaintance had let her know about the institution.
After that, however, she had taken a long time of self-doubt until she actually had contacted the institution.
During the interview, she knew she could have searched online, but she said she wouldn't be able to access only with a phone number posted online.
U3 noted that she had uncertainty about how the process would be like once she contacted those channels, but she had been reluctant to ask further details about support channels to others.
U3 added that, thus, using a chatbot could enhance the sense of safety from the obtained information and the very process of obtaining that information.
U4 initially asked her older sister about where to contact. 
Then, her sister asked her school teachers about that, but they did not know relevant resources and brought her to a generic school counselor who was not necessarily an expert about sexual violence.
All the interviewees commonly reported that they had never used a personalized information system like ours.

\subsubsection{Comfortability}
As survivors might be mentally burdened (U1), discouraged (U2), and self-blaming (U4) while seeking support, they mentioned possible improvements for comfortability regarding sensitivity and accountability.

\bstart{Sensitivity} 
Our interviewees differently perceived the sensitivity of questions in our system.
On the one hand, U3 and U4 said direct and explicit expressions are necessary to pursue clear understanding, so those expressions are inevitable and not overly burdening.
On the other hand, U1 and U2 considered sexual violence survivors' support-seeking as a process of sense-making and empowerment, and they proposed reordering questions in an ascending order of sensitivity.
For example, they felt uncomfortable with some initial questions about sexual penetration in the Emergency Screening phase while they understood why our prototype asked it.
U1 then suggested asking it later after asking about mild physical contacts, which could help survivors gradually identify what they have experienced.
U2 proposed informing users about each question to be asked and offering an option to skip answering, so that survivors can avoid unwanted recall of past memory.
U2 and U3 also said they would want a friendlier speech tone to mitigate mental burdens.

\bstart{Accountability}
As survivors may blame themselves during their sense-making, it is critical to clarify why a system is asking a particular question to help them respond more confidently.
However, the interviewees pointed out that our prototype was limited to rationalize why that question is being asked `at the very moment.'
For example, U4 noted that asking ``did you drink more than usual'' after asking about loss of memory might suggest that survivors could be blamed for drinking, while it is intended for physiological evidence collection.
U1 and U4 suggested clarifying the purposes of asking case details because if a certain detail item does not apply to a user's case, then they might feel their case is denied by the system although the system does not make any judgment based on a single, particular detail.

\subsubsection{Relevancy}
Our interview participants expressed both positive and negative aspects around relevancy of the provided information.
U3 stressed that it is really important to provide as much information as possible because even simple steps like ``calling to a related institution requires information.''
She also mentioned that while she was not particularly interested in how police would be involved in the process, having information outside of her interests helped her to feel comfortable and reliable with a formal process followed by police reporting in addition to satisfying her general curiosity about the process.
U4 further addressed that instructions for both immediate actions (\eg~not washing clothes) and longer term actions (\eg~keeping coherent notes) seemed to be useful to her because survivors can be afraid of unexpected, adverse consequences of what they have already done.
However, attempts to provide more inclusive information regarding emerging needs can make the provided information seem less relevant to the users.
For example, our prototype provides a short paragraph (2-3 sentences) about reporting process. 
U1 and U2 said that they felt difficulty in interpreting such information because their were not interested in it.
U2 also said, after receiving information about counseling, she wanted more details about counseling while U3 said the same content about counseling seem to be sufficient.
We decided not to offer details of counseling program because individual counseling programs highly depend on particular institutions, counselors, clients, and cases, which is a potential source of false expectation in counseling practice (Megan and Jennifer).

\subsubsection{Difficulty}
Although they were able to understand most of the expressions used in our prototype as they have experienced support institutions before, interviewees suggested using easier expressions considering that their earlier understanding had been limited (U1, U4).
U1, U2, and U4 noticed that information about temporally distant steps like juridical processes could be omitted because general public might not have sufficient background knowledge to easily understand that, and because survivors could often be well informed about that once they contact one of the recommended institutions.
Next, in our prototype, we separated questions about oral and non-oral penetrations because instructions for and legal consequences of them are different.
U4 said that survivors in the early stages might have difficulty in distinguishing such similar but different terms.
U4 suggested using multi-select answer options as an alternative to yes-no questions, which may help users to avoid interpreting the same expression multiple times.

In addition, while using an interactive system, a user might feel difficulty in answering to the questions for personalization (\ie~retrieving and providing their own information by answering to the system).
In our interview, users in general evaluated the answering options of our prototype seemed to be easy to use and helpful.
For example, U4 noted that \textit{``I'm not sure'' option} seems to be helpful, considering the hard time she had to have in making sense of information due to the lack of prior knowledge.
U2 noted allowing for choosing multiple purposes of use was useful in the same line, and U4 suggested employing multiple answering options for other sensitive questions.

\subsubsection{Actionable instructions}
Our prototype provides information about possible actions the survivors can take and their possible outcomes, such as how to preserve remaining evidences, advantages of making consistent notes, and an outline of reporting procedures.
Interviewees said that those detailed instructions seemed to be useful.
For example, U2 and P1 (from the focus group interview) noted it is helpful to know that survivors can get medical and counseling support without police engagement.
Although reporting was not her purpose, U1 indicated informing about merits of keeping consistent notes seemed useful because survivors are not likely to hear about it and hence they might not try to make notes at all.
U4 said that ``people know that it is necessary to collect evidence, but they don't know why it is necessary and that statement is really important when there are no evidence to collect.''
U3 said that survivors might feel comfortable by being able to envision what comes after taking actions (\eg~reporting) based on information from trusted sources, which she wished for counseling procedures.
U4 mentioned that information about the need for an adult guardian (regardless of who they should be) and making statements can provide younger survivors with a wider perspective to their situations (based on her experience as an underage survivor).

\section{Personalization Trade-offs and Strategies}
Our interview results imply that users can perceive personalization questions and information with different levels of sensitivity and relevancy, respectively.
This difference in perceived sensitivity and relevancy motivates understanding personalization trade-offs of support-seeking systems for sexual violence survivors for which a single solution or a compact set of rules may exist.
We illustrate personalization trade-offs in terms of understanding users for personalization and providing personalized information, which we learned from both design workshop with practitioners and a user study with target users.
Then, we describe strategies to facilitate reasoning about the trade-offs we identified.
When possible, we include related discussion from prior work to provide further context of our findings.

\subsection{Trade-off in Understanding Users for Personalization}\label{sec:reversed_approach}
To provide personalized information, a system needs to understand users by asking specific questions about their case details, which is also a common practice at the workshop participants' institution.
However, higher specificity may also increase the sensitivity of questions.
When beginning Session 3, Jennifer indicated that asking overly in detail might discourage people from using the system because it may be burdening and raise concerns about data storage, despite anonymity.
During the workshop, Megan said, ``it was really hard to decide how much in detail we should ask [about case details]'' because counselors usually adjust the level of specificity using non-verbal cues (\eg~avoiding questions when a client has an uncomfortable face) during in-person consultation, but non-verbal cues are more limited in a technological system.
While prior work on chatbots~\cite{chaves2019should} shows asking specific questions is useful and desired to provide relatable and profound information, from the perspective of sexual violence survivors, higher sensitivity may discourage survivors from accessing support resources~\cite{quadara2008responding,justice2013}.
We initially framed this problem as \textit{a trade-off between enhancing specificity of information and reducing sensitivity to users}.

To handle the trade-off between specificity and sensitivity, we took a \textit{reversed approach} during the workshop, by which we designed parts of dialogue in a reversed order of user's journey.
Our intended phases of conversation was first asking questions and then providing information, but we designed which set of information to provide first (Session 3), analyzed conditions for each piece of information, and then generated relevant questions in accordance with the conditions (Session 4).
In doing so, we derived a set of rules to determine what to provide: (1) \textit{what is changed by asking?}, (2) \textit{is that change beneficial?}, and (3) \textit{can that change cause harm?}
For example, whether a rape case occurred at work affects the degree of the criminal sentence on the perpetrator (yes to 1).
However, there is little difference in immediate actions for the survivors of workplace and non-workplace rapes, so we decided not to ask about workplace rapes (no to 2).
Moreover, if a system recommends users to take action against the perpetrators of workplace rape, it might cause unexpected substantial damage, such as job security (yes to 3).
On the other hand, we decided to ask about workplace sexual harassment via verbal or behavioral expressions because there are counseling institutions
for workplace sexual harassment that have authority to urge survivors' employers to discipline the offenders properly (yes to 1 and 2).
Compared to the previous prototypes that the researchers created during the sensitizing period, this approach helped us to remove unnecessary sensitive questions.

In the interview, U1 and U2 said some of the initial questions in Emergency Screening phase sounded uncomfortable although they understand why those questions were asked.
To mitigate sensitivity problems, U1 suggested reordering the questions our prototype asks in an ascending order of sensitivity, and U2 proposed allowing for skipping a question.
In a similar line, U1 and U4 recommended enhancing the accountability of questions asked by our prototype to help survivors avoid self-blaming.
Thus, we evolved our initial understanding of the reversed approach to considering the management of sensitivity and specificity as a dyadic penetration process where communicators gradually increases privacy in sharing information~\cite{altman1981}.
Here, privacy means how much private a person thinks a piece of information is (\eg~hobby tends to be less private than medical history).
Practically, designers need to carefully reason about the order of conversation and questions derived using our reversed approach.

\subsection{Trade-off in Providing Personalized Information}\label{sec:inclusive_tailoring}
An information system for sexual violence survivors needs to classify users to provide relevant information to their cases. 
User classification also helps practitioners to provide effective resources.
Megan, an emergency counselor, said that her team categorize their clients for better support.
Vivian, a nurse, also noted that the categorization helps her to optimize solution.
However, the high degree of tailoring might ignore emerging needs of sexual violence survivors that they may develop upon recognizing related information~\cite{skinner2017issas}.
Vivian and Megan indicated that some survivors intend to seek medical support at first and then develop curiosity about police investigation after learning relevant information because those two processes are often interrelated. 
In addition, as U4 pointed out, survivors may not have specific purpose in using a support-seeking system. 
We iterate this as a trade-off between relevancy and inclusiveness.

To address this trade-off, we took an \textbf{inclusive tailoring} approach, which refers to continuously examining how personalization might ignore emerging needs and specify emergent needs.
The underlying motivation is `to prevent unspecified earlier decision from affecting the possibility of users' later actions.'
Inclusive tailoring is not concerned only with how to provide information but how to understand users intention.
Low-level strategies for inclusive tailoring include having a ``not sure'' option, clarifying questions, initial screening, and ordering of information.
First, we attached \textit{an ``I am not sure'' option} to questions regarding the purposes of use and the case details for those who are not able to or do not want to accurately answer to questions due to partial memory loss after crimes~\cite{elliott1995posttraumatic,wakefield1992recovered}, reluctance to answering, or less specified purposes.
If a user chooses this option, more generic information is given with conditional statements.
For example, if a user is not sure whether the perpetrator used drugs, we briefly inform a drug test (only until the test is effective, \ie~72 hours ago).
Hello Cass~\cite{cass} also utilizes this option in asking type of a case.

Second, we included \textit{clarification questions} where the interpretation of a user's response is ambiguous when it must be clear. 
For instance, we added a question asking whether a user was underage at the moment of incident if it is unclear based on the response. 
We clarified this question because it can change the statute of limitations (suspended until they reaches age of 19) and we had not to collect their date of birth.

Third, we added \textit{a screening phase for users' urgent needs} by asking questions regarding the approximate time of occurrence and any incidences of sexual abuse at first.
We defined emergency as ``within less than five days after the incident'' because physiological evidence collection is effective within at most five days, and because severe physical injury, if any, needs immediate care.
When a user is considered to be under an emergency, our prototype provides a list of available institutions and pre-visit instructions, advising continuously using it on their way to a recommended institution (Emergency Information phase). 

Finally, we \textit{ordered parts of provided information} in terms of a user's purposes of use and time of the incident.
We discovered that inclusive tailoring may result in lengthiness, but we could not simply remove some part of the information because a user might have answered inaccurately for the aforementioned reasons.
For example, if a user selects only medical support as their purpose of use, then medical information appears earlier than brief aftereffect consultation information.
When a user's purposes involve medical support but their case occurred quite a long ago, the consultation information appears before the medical information since it is difficult for the user to receive medical care.

While our interview participants found \textit{``I'm not sure'' option} (U4) and \textit{having information outside of her interests} (U3) seem to be useful, attempts to provide more inclusive information regarding emerging needs can make the provided information seem less relevant to the users.
For example, our prototype provides a short paragraph (2-3 sentences) about reporting process. 
U1 and U2 said that they were not interested in it, so they had difficulty in interpreting such information.
U2 also said, after receiving information about counseling, she wanted more details about counseling.
In other words, inclusive tailoring should deal with perceived relevancy of provided information, and an individual support-seeking system should reason about what to prioritize.
Hence, future designs need to address how to satisfy users' emerging needs while reducing mismatched expectation regarding support channels.
\section{Discussion and Future Work}
In this paper, we implemented a prototype dialogue-based interaction system for support-seeking of sexual violence survivors through design workshops with experts and evaluated it with self-identified sexual violence survivors as target users.
The user study participants overall positively evaluated our prototype support-seeking conversation system with suggestions for further improvements. 
By characterizing personalization trade-offs and strategies in designing a dialogue-based information system, we leverage prior findings~\cite{park2021,park2020can,ahn2020chatbot,bauer2019} on designing conversational support systems for survivors. 
Below, we discuss implications of our methods and findings and outline future work toward next-generation support systems for sexual violence survivors. 

\bstart{Method Reflection}
We carefully planned every step of our methods in a way that allows us to gradually develop trust and rapport with our participants. 
As noted earlier, our design workshops with experts helped us not only to derive useful design knowledge but also to build a bridge between the researchers and a network of caregivers, which may have enhanced safety and trustworthiness in participating the interviews.
Through this gradual expansion of participation, we could enrich our perspective and understanding about designing a support system for sexual violence survivors. 
For instance, we first noticed that specificity and sensitivity form an important trade-off in designing a conversational support-seeking system for sexual violence survivors, and we derived and employed the reversed approach to minimize unnecessary disclosure.
After the user interviews with sexual violence survivors, this initial understanding evolved to considering a dialogue system as a dyadic penetration where the temporal order of asking information matters. 
Similarly, after the user interviews we found that our attempts to support both urgent and unspecified needs of survivors indeed translate to challenges in predicting the relevancy of information.

\subsection{Implications for Future Work}
We characterized two personalization trade-offs---sensitivity vs. specificity and relevancy vs. inclusiveness---in designing a conversational information system for sexual violence survivors.
Understanding these trade-offs can inform design processes from defining systems' goal(s) to designing conversation scenarios.
For instance, if a design team plans to deploy a first contact application with lower sensitivity like HelloCass~\cite{cass}, then this system will provide limited specificity but may try to motivate users to look for further tailored information (\eg~linking to other resources).
In contrast, if a system aims to provide case-specific information like ours, then it has to ask sensitive questions for personalization, which some users may find less comfortable, so mitigation methods for such high sensitivity (\eg~justification, conversation flow) are necessary.
For highly personalized systems, a design team may want to support prompt interaction by providing only relevant information.
While prompt interaction is desired for better outcomes, user responses are not always guaranteed to be their full disclosure due to privacy risks~\cite{park2021,freed2018stalker}, so personalization may ignore underspecified needs.
Thus, future work needs to seek methods to balance relevancy and inclusiveness, such as letting users indicate how much in detail they have responded, in addition to our four low-level strategies for inclusive tailoring.
To motivate future designs of informational systems for sexual violence survivors, we outline the following potential next steps in terms of user group scoping, networked support systems, and interface design.

\bstart{Scoping \textit{who} to support}
Sexual violence survivors are not a single user group but have different characteristics in terms of priorities, support-seeking stages, and social contexts.
As our user evaluation implies, survivors have different priorities and preferences.
For example, U3 and U4 stated needs for explicit expressions while U1 and U2 preferred more indirect expressions.
Similarly, U2 sought more information about counseling programs whereas U4 was satisfied with our system for her informational needs.
Next, we designed our system for survivors who have just started seeking support, but those already taking a particular support option may need more specific tools. When it is difficult to make a statement, for instance, a tool that guides how to write a well-formed statement would be more useful than generic tools.
Furthermore, while we adopted guidelines of Korean support organizations to classify user, such classification criteria might be unavailable in some countries. In that case, one may need to examine local context of sexual violence. 
Therefore, future work should explore interactive support-seeking systems by scoping \textit{who} it can support prior to \textit{what} it can do, seeking a deeper understanding of sexual violence survivors as multi-faceted individuals.

\bstart{Networked support systems} 
We encourage future work to reason about a network of support systems for sexual violence survivors with different characteristics and needs.
In practice of support for sexual violence survivors, care-givers (\eg~medical experts, counsellor, \etc) leverage their resource by transferring survivors to a better suited service through their network, which is critical to ensure the victim-centered approach~\cite{csom2018,koss2020,sart2011}.
If we have many support (-seeking) systems in the future, users may find it challenging to choose one, so a network of such systems can help them find best-fitting resources.
For instance, our prototype could be extended to detect and connect users who want to articulate their experience to related services like Spot~\cite{spot}, or at the end of use, our system could provide a link to Gabbie~\cite{gabbie} for those who want to report their case. 
In doing so, future work may need to pursue a deeper understanding of users' journey in seeking support using technology, extending prior work on their use of social media in seeking support~\cite{razi2020let} and how their circumstances impact support-seeking~\cite{freed2018stalker}.

\bstart{Future system and interface design}
To cope with personalization trade-offs, we derived two design approaches (reversed approach and inclusive tailoring) in the context of a close-ended dialogue system.
Future work could leverage these approaches for other formats like Web questionnaire or open-ended dialogue.
High-level implication of our reversed approach is that concrete definition of feedback to users needs to precede designing interface for user input, and that of inclusive tailoring is that designers need to consider how to account underspecified and/or emerging needs.
A reversed approach for a Web questionnaire, for example, could outline feedback sections first, derive conditions for each section, and make question items for those conditions.
By applying inclusive tailoring, An open-ended chatbot could employ a rule-based, probabilistic, or hybrid inference model to screen urgency and detect underspecified needs.

Our approaches have downstream impact to back-end components, such as information management and knowledge representation.
For examples, it is critical to update support-seeking systems with latest information (\eg~new resources, deprecated services), which can translate to a problem of how to manage and represent knowledge base for such systems.
Ideally, a knowledge base should allow both human practitioners and the system to easily and immediately update related information.
Even though a practitioner can easily request an update, if knowledge representation is complex (\eg hard-coded scripts), that it will be more tedious to implement the update~\cite{hicks2001dynamic}.
Our prototype uses logical operators ($=$, $\&$, \etc) to represent conditions for each piece of conversation to simplify both encoding and compiling those conditions.
Future work could consider other formalization techniques including truth maintenance systems~\cite{doyle1979truth}, Answer Set Programming~\cite{Gerhard2011}, Planning Domain Definition Language~\cite{mcdermott20001998}, as well as machine learning (ML)~\cite{bauer2019,park2021}.
Future work with probabilistic ML modeling (\eg~\cite{bauer2019,park2021}) should be clear about (but not limited to) data collection and processing, performance evaluation, and optimization goals, as raised by recent works on model reporting~\cite{mitchell2019modelcards,gebru2020datasheets}.

\subsection{Limitations}
We used a conversational information system that can be deployed online as a test bed.
This research was done in a country with high digital literacy (Republic of Korea) where most people have access to public or private internet. 
Thus, future work should explore designs for regions with limited access to internet or low digital literacy.
We also note that while we intend this system to be used to inform sexual violence survivors about access to professional support resources, some organizations might appropriate our findings to replace professional care-givers with such systems.
However, we emphasize that we have never sought systems for replacing human care-givers, but for connecting survivors to those human care-givers, and hence our findings are not applicable for such replacing systems.

\section{Conclusion}
Through four design workshop sessions with three domain experts, we implemented a prototype conversational system for support-seeking of sexual violence survivors.
After qualitatively evaluating it with four self-identified sexual violence survivors, we characterize personalization trade-offs in designing such a dialogue-based system and strategies to handle those trade-offs. 
In particular, we utilized the reversed approach and inclusive tailoring to deal with personalization trade-offs around sensitivity and inclusiveness.
The scope of our work includes informational support, adult users, widely known crime types, dialogue-based interaction.
While we attempted to incorporate multiple forms of participation, extensive user participation and in-practice deployment are necessary to deepen our findings.
Thus, future design studies should explore different aspects of sexual assault through extended engagement with users and real world deployment, which will reveal further unique, unexpected design challenges.

\begin{acks}
Special thanks to Inha Cha. 
This work was supported by the Ministry of Education of the Republic of Korea and the National Research Foundation of Korea (NRF-2021S1A5B8096358).
\end{acks}

\bibliographystyle{ACM-Reference-Format}
\bibliography{main}

\end{document}